\documentstyle[preprint,aps]{revtex}
\tighten
\begin{document}
\draft
\title{Electronic structure of Fibonacci Si $\delta$-doped GaAs}
\author{F.\ Dom\'{\i}nguez-Adame, E.\ Maci\'a$^*$, and B. M\'endez}
\address{Departamento de F\'{\i}sica de Materiales,
Facultad de F\'{\i}sicas, Universidad Complutense,
E-28040 Madrid, Spain}

\date{\today}
\maketitle
\begin{abstract}

We study the electronic structure of a new type of Fibonacci
superlattice based on Si $\delta$-doped GaAs.  Assuming that
$\delta$-doped layers are equally spaced, quasiperiodicity is introduced
by selecting two different donor concentrations and arranging them
according to the Fibonacci series along the growth direction.  The
one-electron potential due to $\delta$-doping is obtained by means of
the Thomas-Fermi approach.  The resulting energy spectrum is then found
by solving the corresponding effective-mass wave equation.  We find that
a self-similar spectrum can be seen in the band structure.  Electronic
transport properties of samples are also discussed and related to the
degree of spatial localization of electronic envelope-functions.

\end{abstract}

\narrowtext

\section{Introduction}

Following the first fabrication of quasiperiodic semiconductor
superlattices \cite{Merlin,Todd}, there has been a increasing interest
in the study of one-dimensional systems describing quasiperiodic
structures.  From the very beginning, most researchers have considered
the Fibonacci sequence as a typical example of a quasiperiodic system
\cite{Koh,Ostlund}, and several characteristic properties of Fibonacci
systems have been reported during the last decade.  Thus, it is now well
established that Fibonacci lattices exhibit highly fragmented electron
and phonon spectra with a hierarchy of splitting subbands displaying
self-similar patterns \cite{Laruelle}, and their corresponding
electronic density of states shows spiky features \cite{Hirose}.  This
exotic electronic spectrum strongly influences electron propagation
\cite{Katsumoto,Angel} and dc conductance through the system, even at
finite temperature \cite{Enrique}.  Furthermore, electronic wave
functions are neither extended, in the Bloch sense, nor exponentially
localised; instead, they are critical in Fibonacci lattices
\cite{Chakrabarti}.

All these striking results, however, have been obtained for two
particular kinds of model Hamiltonians, namely tight-binding models
(either diagonal, off-diagonal or both) and Kronig-Penney models.
Therefore, the question as to whether those features purported so far
as characteristic of Fibonacci order are extensible to more realistic
Fibonacci Hamiltonians becomes very appealing from a theoretical point
of view.  On the other side, since it was realized that Fibonacci
superlattices (FSLs) preserve its quasiperiodic order even if
substantial growth fluctuations in the sequential deposition of layers
is allowed for \cite{Todd}, a considerable interest on the transport
properties of quasiperiodic semiconductor heterostructures has emerged
in the experimental field.  However, FSLs studied up to date are
quantum-well GaAs/AlAs based ones.  This shortcoming does not seem
necessary from an experimental viewpoint, since computer controlled
doping deposition techniques are currently able to construct a wide
variety of superlattice potential profiles.  In particular,
epitaxial-growth techniques allow to prepare $\delta$-doped
semiconductor structures, in which a sheet of donors is localized within
a few monolayers of the crystal.  These impurity atoms supply electrons
and give rise to strong confinement by the resulting one-electron
potential.

The main purpose of this letter is to show that {\em distinctive}
features, previously obtained for simpler Fibonacci Hamiltonians, also
appear in more complex and realistic systems, hence strongly suggesting
that those features can be a {\em universal} fingerprint of
one-dimensional Fibonacci systems.  To this end, in the present work we
propose a new type of FSL based on quasiperiodically Si $\delta$-doped
GaAs.  We study the resulting electronic structure within the
envelope-function and effective-mass approximation.  The one-electron
potential is found by solving the Thomas-Fermi equation.  We find that
the electronic spectrum is highly fragmented and shows self-similar
patterns which become very sensitive to the doping ratio concentration.
The spectral type of our system is analysed by means of
bandwidth-scaling techniques suggesting a underlying singular continuous
character.  Finally, transport properties of the FSL at zero temperature
are discussed in the context of the Landauer formula and related to the
possible critical nature of the electronic states.

\section{The Model}

The system we study in this work is a FSL made of Si $\delta$-doped
GaAs.  In general, a FSL of order $N$ is generated from two basic units
A and B by successive applications of the inflation rule
A~$\rightarrow$~AB and B~$\rightarrow$~A yielding a sequence of the form
ABAABABA \ldots\ This sequence comprises $F_{N-1}$ elements A and
$F_{N-2}$ elements B, $F_l$ being the {\em l\/}th Fibonacci number given
by the recurrent law $F_l=F_{l-1}+F_{l-2}$ with the initial values
$F_0=F_1=1$.  In the present model we take each unit (A or B) as a slab
of GaAs of thickness $a=200\,$\AA\ with a Si $\delta$-doped layer
embedded in its center.  The quasiperiodicity is then introduced by
allowing the doping layers to take on two possible values of donor
concentration, and arranging them according to the Fibonacci sequence.
Each doping layer consists of a continuous positive slab of thickness
$d=50\,$\AA\ with either $N_D^{(A)}$ or $N_D^{(B)}$ ionized donors per
unit area.  Hereafter we fix $N_D^{(A)}=5\times 10^{12}\,$cm$^{-2}$, and
allow $N_D^{(B)}$ to vary from $2.0$ up to $6.5\times
10^{12}\,$cm$^{-2}$ in order to study the influence of the doping ratio
$r \equiv N_D^{(B)}/N_D^{(A)}$ onto the resulting electronic
spectrum.  Finally, we assume that there exists a uniform p-type
background doping with $N_A=10^{15}\,$cm$^{-3}$ acceptors per unit
volume. This range of parameters corresponds to samples which could be
easily grown by molecular beam epitaxy techniques now available.

First we compute the one-electron potential $V_j(x)$ in each basic unit
(here $j$ denotes A or B), where $x$ indicates the spatial coordinate
along the growth direction.  This we accomplish by means of the
Thomas-Fermi (TF) semiclassical model.  It is known that the TF
formulation is equivalent to the self-consistent formulation in a wide
range of doping concentrations \cite{Ioriatti}, and it has been
successfully applied to periodically Si $\delta$-doped GaAs SLs
\cite{Egues,nosotros1,nosotros2}.  Hence we can confidently adopt this
approach to calculate the one-electron potential due to the layer of
ionized donors.  Choosing the middle of the $\delta$-doped layer as the
origin of coordinates, the nonlinear TF differential equation reads
\begin{eqnarray}
\frac{d^2V_j(x)}{dx^2}&=&-\,{8\over
3\pi} \left[\epsilon_F-V_j(x)\right]^{3/2}
+{8\pi \over d}N_D^{(j)} \theta \left( {d\over 2}+x\right )
\theta \left({d\over2}-x \right) \nonumber \\
&-&\, 8\pi N_A \theta \left( {a\over 2}+x \right)
\theta \left({a\over 2}-x \right), \hspace{1cm} j=\mbox{A,\ B},
\label{TF}
\end{eqnarray}
where $\epsilon_F$ denotes the Fermi energy.  The value of the Fermi
energy cannot be computed self-consistently in the TF formulation, so it
must be regarded as a parameter.  Thus we fix the origin of energies so
$\epsilon_F=0$ henceafter.  Distances and energies are scaled in units
of the effective Bohr radius a$^*$ ($=100\,$\AA) and effective Rydberg
Ry$^*$ ($=5.8\,$meV), respectively.  The boundary conditions for this
equation are those of the superlattice given by \cite{Egues}
$V^\prime_j(0) = V^\prime_j(a/2) = 0$.  The resulting potential becomes
deeper in the gap on incresing donor concentration.

We assume the validity of the effective-mass approximation, and we take
isotropic and parabolic conduction band in the growth direction.  This
approach works well in some direct-gap compounds, as it is the case of
GaAs.  Once the potential $V_j(x)$ in each basic unit is found from
(\ref{TF}), the electron envelope-function and energy values can be
obtained from the following one-dimensional Schr\"odinger equation
\begin{equation}
-\,\frac{d^2\psi (x)}{dx^2}+ V(x) \psi (x) = E \psi (x),
\label{Schrodinger}
\end{equation}
where $V(x)$ is the potential of the whole FSL. We assume that this
potential at any point of the system is given by the one-electron
potential $V_j(x)$ we have just computed in each basic unit, where $V_A$
and $V_B$ are arranged according to the Fibonacci series. Therefore,
$V(x)$ is simple a piecewise potential with $F_N$ regions.

\section{Numerical analysis}

Since there is no analytical expression for the potential $V_j(x)$,
computations must rely on numerical procedures.  To this end, we
divide the each basic unit of the FSL in a grid of points $\{ x_k=kh\}$,
where $h=a/n$ is the integration step and $n$ is the number of grid
points in each basic unit.  In our calculations we have taken $n=400$,
which is more than enough to obtain very accurate results.  The
discretized form of the Schr\"odinger equation (\ref{Schrodinger}) may
be cast in the matrix form
\begin{equation}
\left( \begin{array}{c} \psi (x_{k+1}) \\ \psi (x_k)\end{array} \right)
= \left( \begin{array}{cc} \alpha_k & -1 \\
1&0 \end{array} \right)
\left( \begin{array}{c} \psi (x_k) \\ \psi (x_{k-1})\end{array} \right)
\equiv P_k
\left( \begin{array}{c} \psi (x_k) \\ \psi (x_{k-1})\end{array}\right),
\label{Pn}
\end{equation}
where we have defined $\alpha_k\equiv 2+h^2(V(x_k)-E)$ for brevity.
Iterating this equation one obtains
\begin{equation}
\left( \begin{array}{c} \psi (x_{N_{SL}+1}) \\ \psi (x_{N_{SL}})\end{array}
\right) = P_{N_{SL}}\, \cdots \, P_0
\left( \begin{array}{c} \psi (x_0) \\ \psi (x_{-1})\end{array} \right)
\equiv T(N_{SL})
\left( \begin{array}{c} \psi (x_0) \\ \psi (x_{-1})\end{array} \right).
\label{Tn}
\end{equation}
$T(N_{SL})$ is the transfer matrix of the FSL and $N_{SL}=nF_N$ is the
number of grid points in the whole structure. $T(N_{SL})$ is real and
relates the envelope-function at both edges of the structure.  Taking
into account that $T(k)=P_{k}\, T(k-1)$ and $T(0)=P_0$ we find the
following recurrence relations involving only real parameters
\begin{eqnarray}
T_{11}(k) &=& \alpha_{k} T_{11}(k-1)- T_{11}(k-2),  \nonumber \\
T_{12}(k) &=& \alpha_{k} T_{12}(k-1)- T_{12}(k-2),  \nonumber \\
T_{21}(k) &=& T_{11}(k-1),  \nonumber  \\
T_{22}(k) &=& T_{12}(k-1), \hspace{1 true cm} k=1,2\, \cdots N_{SL}.
\label{recurr}
\end{eqnarray}
These equations must be supplemented with the initial conditions
$T_{ij}(-1)=\delta_{ij}$, $T_{11}(0)= \alpha_0$, $T_{12}(0)=-1$,
$T_{21}(0)=1$ and $T_{22}(0)=0$.  The resulting allowed energies can be
found by imposing periodic boundary at both edges of the system with
$F_N$ layers arranged according to the Fibonacci sequence. Once we have
obtained the transfer matrix for the whole FSL, the resulting
energies for which the absolute value of the trace of the
transfer-matrix corresponding to the whole FSL is smaller (larger) than
$2$ are allowed (forbbiden) \cite{Ziman}.

\section{Results and discussions}

In order to properly discuss the novel features arising from
quasiperiodicity in our system, it is convenient to give a brief account
of the electronic spectrum associated to this model when all $\delta$
layers have the same donor concentration. In this case, which
corresponds to a usual periodic superlattice, resonant coupling between
identical states of neighbouring layers leads to the formation of
minibands of finite width \cite{Egues,nosotros1,nosotros2}, related
to extended (Bloch type) electronic states. The resulting miniband
structure and the dispersion relation inside allowed minibands $E(K)$,
$K$ being the momentum perpendicular to the layers, can be evaluated
using the transfer-matrix approach as well. Since Bloch theorem must be
satisfied in the periodic case, the dispersion relation is found to be
(see Ref.~\onlinecite{jpa} for details)
\begin{equation}
\cos (K a) = {1\over 2} \mbox{Tr} [T(n)].
\label{dispersion}
\end{equation}
Notice that the required time-reversal symmetry of the dispersion
relation $E(-K) =E(K)$ is conserved.  As an illustration, we show in
Figure ~\ref{fig1} the miniband structure for a particular realization
corresponding to a donor concentration of $5\times 10^{12}\,$cm$^{-2}$.
Note that the lower miniband is almost nondispersive since its width is
rather small (about $0.4\,$Ry$^*$).  This suggests that the ground state
of the Thomas-Fermi potential is only weakly coupled to its neighbouring
wells.  On the contrary, the second miniband is clearly dispersive, as
seen in Fig.~\ref{fig1}, and its width amounts $4.7\,$Ry$^*$.  The third
miniband is even wider and it crosses the Fermi level.

Now, we consider the most prominent features of the resulting electronic
structure when quasiperiodicity is introduced.  A schematic diagram of
the FSL potential for $F_5=8$ is presented in Fig.~\ref{new}.  From a
mathematical point of view, one of the most characteristic properties of
electronic spectra in Fibonacci systems is its highly fragmented,
Cantor-like nature.  We have confirmed this fragmentation in our FSL
even when deviation from perfect periodicity is actually small, in other
words, when the ratio $r$ is close to unity.  In fact, we have found
that each miniband of the periodic SL, shown in Fig.~\ref{fig1}, splits
in several sub-minibands, that is, small gaps appear.  The origin of
these small inner gaps are directly related to the loss of long-range
quantum coherence of the electrons, as the potential inside basic units,
A and B, becomes different.  Results corresponding to the fragmentation
of the second miniband are shown in Fig.~\ref{fig2} as a function of the
Fibonacci order $N$.  We have mainly focused on this second miniband
since the first one is almost nondispersive and investigation of the
fragmentation process would require very tiny energy steps and then is
rather time consuming.  Only short approximants of the FSL are displayed
in Fig.~\ref{fig2} since on increasing $N$ the spectrum becomes so
fragmented that it is difficult to observe minor features in the plot.
However, we have carefully analyzed FSLs spectra up to order $N=12$ (233
layers) and we have confirmed that the number of sub-minibands composing
the whole spectrum is exactly $F_N$, i.e. the number of basic units
forming the superlattice.

As we mentioned in the Introduction, another characteristic feature of
Fibonacci systems is the self-similar pattern exhibited by their
corresponding spectra.  This self-similarity has been widely
investigated within the tight-binding approximation while much less work
has been devoted to quantum-well superlattices.  Our results show that
self-similar spectra are also obtained in Fibonacci Si $\delta$-doped
GaAs, as shown in Fig.~\ref{fig3}.  It is clear that the whole
electronic spectrum for a short approximant ($F_4=5$ in this case) is
mapped onto a small portion of the spectrum of a higher approximant
($F_7=21$ in Fig.~\ref{fig3}).  This is a consequence of how the FSL is
constructed, based on a deterministic substitution sequence
\cite{lamadredeKomoto}.

A third characteristic of Fibonacci systems concerns their spectral
type, which results to be singular continuous.  In order to estimate
the spectral type associated to our model Hamiltonian, we have
calculated the normalized equivalent bandwidth $S$, defined as the ratio
between the sum of all allowed sub-minibands and the width of the second
miniband in the periodic $A$ SL ($4.7\,$Ry$^*$).  As can be expected
from the Cantor-like nature of Fibonaccian spectra, $S$ vanishes as the
system size grows.  Furthermore, we have obtained that the normalized
equivalent bandwidth decreases according to a power law of the form
$S=F_N^{-\beta}$ with $\beta \sim 0.1$, as seen in Fig.~\ref{figx}.
According with earlier works \cite{lamadredeKomoto}, such a behaviour is
characteristic of a singular continuous spectrum for which all the
envelope-functions are critical, i.e., regarding localization properties
the functions are neither exponentially localized nor extended.

The richness in structure displayed by the electronic spectrum should be
reflected, to some extent, in its transport properties \cite{Enrique}.
We have evaluated the electrical resistance at zero temperature, $\rho$,
using the well-known dimensionless single-channel Landauer formula
\cite{Landauer}: $\rho=R(\epsilon_F)/T(\epsilon_F)$, where $R$ and $T$
denote the reflection and transmission coefficients, respectively.  The
transmission coefficient can be obtained in a straightforward manner in
the transfer-matrix formalism \cite{jpa}.  A typical example of the
obtained results is shown in Fig.~\ref{fig4}(a) for $F_{11}=144$.  We
observe that the Landauer resistance exhibits a highly fragmented
structure displaying dramatic fluctuations under minor variations of the
ratio $r$.  The practical implications of such behaviour should be
clear: One can select almost any desired value of the resistance of the
sample by choosing properly the suitable ratio $r$.  According to
previous works \cite{Enrique}, on increasing temperature interesting
behaviours of the resistance can be expected since the detailed
structure of the energy spectrum naturally determines the finer details
of the resistance pattern at finite temperature.

We have also investigated the spatial extent of the electron
envelope-functions at the Fermi level.  This we accomplished by means of
the inverse participation ratio (IPR), as defined, for instance, in
Ref.~\onlinecite{Hirose}
\begin{equation}
\mbox{IPR}=\frac{\sum_k\ |\psi (x_k)|^4}{\left(\sum_k\ |\psi
(x_k)|^2\right)^2},
\label{IPR}
\end{equation}
where the index $k$ runs over grid points of the whole FSL. The IPR
gives an estimation of the volume occupied by the electron
envelope-function: The smaller the IPR, the more extended the electron
state.  Although a more detailed analysis of the envelope-functions is
required to determine the exact nature of the wave function
(multifractal analysis), for an understanding of the resistance
behaviour the IPR is sufficient.  Figure~\ref{fig4}(b) shows the IPR for
$F_{11}=144$ as a function of the ratio $r$.  Note that the minimum
value is reached when this ratio becomes unity, i.e. in the periodic
superlattice.  The IPR also present dramatic fluctuations under minor
variation of the donor concentration $N_D^{(B)}$.  A comparison of both
Fig.~\ref{fig4}(a) and Fig.~\ref{fig4}(b) reveals that the IPR is larger
(electronic states are more localized) whenever the resistance
increases, indicating that the spatial extent of the envelope-functions
controls the electrical transport of the sample.

\section{Conclusions}

We have proposed a new type of quasiperiodic (Fibonacci) superlattice
heterostructure based on Si $\delta$-doped GaAs.  The corresponding
electronic spectrum shows a highly fragmented, self-similar nature
resembling that found for simpler tight-binding models.  The spectral
type of our model Hamiltonian, obtained from bandwidth-scaling
considerations, indicates that it is singular continuous in the
thermodynamical limit, in agreement with the current opinion of the
mathematical community, supporting the conjecture that the spectral type
for almost all substitution sequences should be singular continuous
\cite{Ghez}.  The spectrum structure is very sensitive to the doping
difference between the basic units, and this fact significantly affects
the transport properties of the sample at zero temperature.  This
interesting result suggests the possibility of a certain degree of
``engineering" of transport properties during superlattice growth by a
proper selection of the corresponding doping sequences.  Finally, we are
able to relate resistance fluctuations to the nature of the electronic
wave function, through the inverse participation ratio.  The obtained
relationship suggests that the overall conductance of the superlattice
is directly connected with the decay rate of the electron wave function
along the sample.  In this sense, it would be convenient to perform a
complete analysis of the wave function espatial distribution by means of
the multifractal formalism.  Work in this direction is currently in
progress and we expect to report on it elsewhere.

\acknowledgments
The authors thank A.\ S\'anchez for a critical reading of the
manuscript. This work has been partially supported by Univeridad
Complutense under project PR161/93-4811.

\begin{figure}
\caption{Miniband structure for periodically Si $\delta$-doped GaAs with
a donor concentration $5\times 10^{12}\,$cm$^{-2}$.}
\label{fig1}
\end{figure}

\begin{figure}
\caption{Schematic diagram of the one-electron potential for a Fibonacci
Si $\delta$-doped GaAs with $N_D^{(A)}=5\times 10^{12}\,$cm$^{-2}$ and
$N_D^{(B)}=4.5\times 10^{12}\,$cm$^{-2}$ with $N=F_5=8$ layers.}
\label{new}
\end{figure}

\begin{figure}
\caption{Allowed sub-minibands as a function of the Fibonacci order $N$,
for a Fibonacci Si $\delta$-doped GaAs with $N_D^{(A)}=5\times
10^{12}\,$cm$^{-2}$ and $N_D^{(B)}=4.5\times 10^{12}\,$cm$^{-2}$. The
number of sub-minibands is $F_N$ for each order $N$.}
\label{fig2}
\end{figure}

\begin{figure}
\caption{Self-similar spectrum of a Fibonacci Si $\delta$-doped GaAs
with $N_D^{(A)}=5\times 10^{12}\,$cm$^{-2}$ and $N_D^{(B)}=4.5\times
10^{12}\,$cm$^{-2}$.  Left plot shows the whole spectrum of a FSL of
order $N=4$ whereas right plot shows a detail of the spectrum of the FSL
of order $N=7$.}
\label{fig3}
\end{figure}

\begin{figure}
\caption{Log-log plot of the normalized equivalent bandwidth $S$ as a
function of the number of $\delta$-doped layers $F_N$ for a Fibonacci Si
$\delta$-doped GaAs with $N_D^{(A)}=5\times 10^{12}\,$cm$^{-2}$ and
$N_D^{(B)}=4.5\times 10^{12}\,$cm$^{-2}$.}
\label{figx}
\end{figure}

\begin{figure}
\caption{(a) Landauer resistance at the Fermi energy and (b) IPR as a
function of the ratio $N_D^{(B)}/N_D^{(A)}$ for a Fibonacci Si
$\delta$-doped GaAs with $N_D^{(A)}=5\times 10^{12}\,$cm$^{-2}$ and
$F_{11}=144$ layers.}
\label{fig4}
\end{figure}

\end{document}